# Generating non-diffracting bottle beams with a flat multi-level diffractive lens

**Andra Naresh Kumar Reddy**[1,2,3,4, a], **Srinivasa Rao Allam**[2,3], **Ashish Tiwari**[5], **Vishwa Pal**[6], **Tina M. Heyward**[7], **Rajesh Menon**[7,8], **and Takashige Omatsu**[2,3,9]

[1]Quantlight and High Harmonics Lab Pvt. Ltd., Door No. 3-7-400/PGRC/304, Survey No. 30/P, Street No. 16, Nalanda Nagar, Hyderguda, Hyderabad 500048, India
[2]Graduate School of Engineering, Chiba University, Chiba 263-8522, Japan
[3]Molecular Chirality Research Centre, Chiba University, Chiba 263-8522, Japan
[4]Laboratory of Nonlinear Optics, Institute of Astronomy, University of Latvia, Jelgavas 3, LV-1004 Riga, Latvia
[5]Laser Centre, Institute of Physical Chemistry, Polish Academy of Sciences, 01-224 Warsaw, Poland
[6]Department of Physics, Indian Institute of Technology Ropar, Rupnagar 140001, Punjab, India
[7]Department of Electrical and Computer Engineering, University of Utah, Salt Lake City, Utah 84112, USA
[8]Oblate Optics, Inc., San Diego, California 92130, USA
[9]Department of Electrophysics, National Yang Ming Chiao Tung University, Hsinchu 300093, Taiwan

[a] **Author to whom correspondence should be addressed:** naarereddy@gmail.com

**ABSTRACT**

We introduce a novel method for creating a high-quality, sharply defined, non-diffracting optical bottle beam by focusing a Bessel beam propagating through a flat multi-level diffractive lens (MDL). This study highlights the impact of the MDL illuminated by a Bessel beam with suppressed sidelobes generated from a binary axicon. The resulting Bessel bottle beam exhibits a series of low- or zero-intensity zones interleaved with high-intensity regions, with variable periods ranging from 0.2 to 1.36 mm along the beam propagation. The transverse intensity profiles of these regions remain shape-invariant over long distances in free space, and thereby the non-diffracting range of the micron-sized optical bottle beam exceeds 5 cm. We also observe that the far-field output from the MDL illuminated by a Bessel beam offers advantages over conventional focusing lenses. Furthermore, this technique can operate on ultrafast timescales (from pico- to femtoseconds) due to the high damage thresholds of the binary axicon and MDL, enabling the generation of high-power optical bottle beams. Ultimately, our experimental approach paves the way for various applications, including high-resolution biological imaging in turbid media, particle manipulation, micromachining, and harmonic generation, leveraging the spatial landscape of the optical bottle beam.

## I. INTRODUCTION

Most laser sources produce Gaussian beams that diverge as they propagate, increasing in size. Tightly focused beams diverge more during free-space propagation. Since the evolution of structured laser beams, significant progress has been made in generating a variety of structured light beams with propagation features by controlling amplitude, phase, and polarisation distributions that differ from the propagation properties of conventional Gaussian beams [1-3], along with their prospective applications [4-10]. The evolution of initial methods involved producing modes such as Laguerre-Gaussian (LG), Hermite-Gaussian (HG), and Ince-Gaussian (IG) by interacting with diffractive optical elements or holograms, resulting in laser beam shapes with benefits in science and engineering [1-3].

Furthermore, among several structured light beams, Bessel beams have garnered considerable interest and a growing demand in the scientific community [12, 13]. They are generated through the self-interference of laser beams as they propagate through space [12]. Notably, the free-space propagation and non-diffracting characteristics of Bessel beams present valuable distinctions compared to the behaviour of eigen and non-eigen modes of generalised Gaussian modes. However, LG beams and Bessel beams have received significant importance over other structured light beams, demonstrating their concrete role in applications due to helicity in the wavefront or orbital angular momentum (OAM) of LG beams [1] and the non-diffracting, self-healing properties of Bessel beams [12, 14-15]. Specifically, conventional Bessel beams maintain their transverse shape as they propagate through free space and comprise exact solutions to the Helmholtz equation in circular cylindrical coordinates [16-17]. However, their large transverse distributions with numerous sidelobes, as well as the complex ring structure, limit their wide applicability in scientific and industrial applications. Additionally, more light beams, whose transverse shape remains invariant over a certain propagation distance in space, such as needle-shaped and pin-like light beams, have demonstrated their potential advantages in optical coherence tomography [18] and free-space optical communication [19], respectively. These experimental considerations demand shaped laser beams with unlimited energy to maintain non-diffracting properties during free-space propagation. The simplest way to generate focused, non-diffracting beams is to use a circular aperture or an axicon [20]. Non-diffracting beams affected by spherical aberrations from the axilens can be controlled with custom adaptive optics to extend the depth of focus (DOF) [21]. Studies have demonstrated optical beams and imaging systems with extended DOF using amplitude-phase and power-phase apodisations, logarithmic axicons, axilens, phase pupil masks, optical diffusers, and iterative algorithms [22- 27]. Among these, Bessel beams, including zero- and higher-order types, are widely studied for their interactions with surfaces and matter [12, 28-33]. However, these methods often have limitations, such as complex setups that depend on polarisation, beam quality, and the wavelength of the illuminating Gaussian beam.

Interest in coherent optical bottle beams has increased significantly due to their multiple low- or zero-intensity regions surrounded by high-intensity areas during propagation [34, 35], making them valuable in both research and technology [36]. Bottle beams have been created using lasers with various modes, such as Gaussian, LG, Airy, and Bessel [37, 1, 38, 34, 39]. Three-dimensional Bessel optical bottles exhibit unique properties, such as self-healing and non-diffraction [40], which enable their use in micromanipulation across various fields, including chemistry, biology, and medicine [41]. Bessel bottle beams offer applications that include dark optical trapping [42], trapping cooled atoms [43], and microsphere separation and imaging in super-resolution microscopy [44]. Several methods validate optical bottle beams using the interference of Bessel beams, while axicons, slits, and digital devices (SLM, DMD) [12, 34, 45-47] are employed to generate Bessel beams. Among the various techniques, axicons are particularly popular because they can withstand high laser power and operate effectively across a wide range of wavelengths. In this context, a compact setup with two axicons can produce high-power Bessel bottle beams, although aligning the two axicons is difficult [40]. Furthermore, a simpler approach is explored by utilising a single binary axicon created via e-beam lithography, which reduces complexity in the experimental generation of Bessel bottle beams [48]. In particular, it is essential to explore the compact method that produces high-quality, micron-sized bottles with an extended depth of focus, which is vital for applications. Further, a simpler and cost-effective method enhances applicability and adoptability.

In the present work, we propose a simple method for generating a high-quality, new class of non-diffracting optical bottle beam with three-dimensional dark cores exhibiting sharp features, which relies on the propagation of a Bessel beam through a flat multi-level diffractive lens (MDL). The MDL was created using inverse design and fabricated with a greyscale photolithography technique [49-53]. The MDL focuses the incoming Bessel beam generated by the binary axicon at a specific working distance in free space. At the same time, the depth of focus of the resultant Bessel optical bottle beam has been extended several times during free-space propagation. Specifically, as the Bessel input beam propagates along the $z$-direction through MDL, starting from the focal plane ($z = 0$), the input beam experiences a dynamic focusing effect caused by the MDL phase, resulting in strong interference of two different wave vectors. This results in the formation of a micron-sized optical bottle beam, characterized by a series of alternating zero-intensity and high-intensity regions along the beam propagation. The far-field intensity distribution of the Bessel beam generated by a flat MDL with a focal length of $f \approx 20$ cm provides useful insights compared to the far-field intensity distribution produced by a traditional focusing lens with a similar focal length ($f = 20$ cm). Furthermore, it is feasible to produce a high-energy optical bottle beam with identical quality and characteristics by using intense femtosecond pulses. We can then investigate nonlinear frequency conversion for the efficient generation of harmonic Bessel bottle beams in the UV and XUV spectral regimes. Note that the MDL used in the experimental procedure is a thin and flat diffractive optical element that transmits the illuminated Bessel beam with a transmission efficiency of approximately 90%. Thus, combining the binary axicon with the MDL produces a high-quality, sharply focused optical bottle beam with an elongated depth of focus. This promising effect offers substantial benefits in various applications, particularly in micromachining, microfabrication, optical trapping, and biological imaging.

## II. EXTENDED DEPTH-OF-FOCUS WITH A FLAT MULTI-LEVEL DIFFRACTIVE LENS

The MDL exemplifies advanced focusing optics developed through the innovative inverse design methodology. This technique permits the precise tailoring of a field pattern specific to the lens plane, thereby enabling the generation of the desired field pattern at the focal plane upon illumination with light [49-51]. Notably, this flat diffractive lens is comparable to a simulated hologram, where the phase retardation is meticulously chosen through a numerical inverse design process. This advancement demonstrates how contemporary design strategies can optimise optical performance and broaden the horizons of lens technology. The foundational principle underlying this inverse design was previously documented [52].

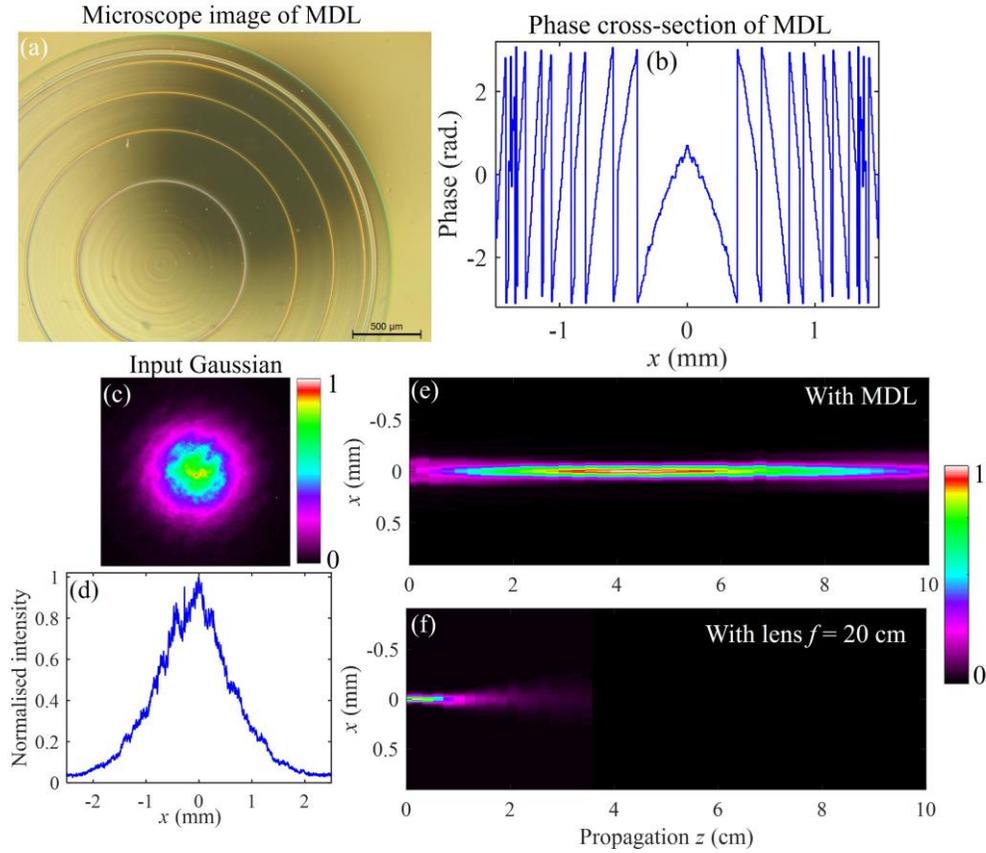

**Fig. 1.** (a) Optical microscope image of MDL and its phase profile given in (b). (c) Spatial intensity distribution ($xy$ - plane) and (d) line profile (along $x$ - axis) of the Gaussian beam used in our experiment. The intensity distribution of a Gaussian beam in its longitudinal beam cross-section ($xz$-plane) while it is propagating via (e) MDL of working distance ~20 cm and (f) through a conventional focusing lens of focal length, $f$ = 20 cm.

The focusing characteristics of the MDL differ from those of traditional lenses, which depend on a stringent parabolic phase to convert incident plane waves into spherical waves. This strict phase paradigm does not rectify aberrations in optical imaging systems where image intensity at the focal plane is paramount, whereas phase considerations are secondary. Consequently, the phase at the focal plane, regarded as a free parameter, may be advantageous in designing a flat diffractive lens that achieves achromaticity and an extended depth of focus [50].

The present MDL design was modeled using Fresnel diffraction theory. A nonlinear optimization method based on the modified direct-binary-search algorithm was employed to determine the optimal heights of the circular zones, thereby maximizing focusing efficiency [51, 53, 56]. Typically, increasing the maximum height of the zones and decreasing their width to a smaller value results in a higher focusing efficiency, which is computed for each focal plane within the desired focal range [51, 53]. The focal range is defined as the span of distances measured from the MDL over which the light is intended to be focused [$f_{max} - f_{min}$]. Since the MDL is polarization insensitive, and with unpolarized light input, the beam propagation begins at the lens plane and continues all the way to the observation plane, covering the entire path in z range, where $z \in (f_{min}, f_{max})$ [56].

Thus, the enrichment in depth of focus (DOF) beyond the diffraction limit is provided by [56]

$$E_{\text{DOF}} = \frac{f_{max} - f_{min}}{\frac{\lambda}{\max(NA)^2}} \quad (1), \qquad \text{since NA} = \sin\left(\tan^{-1}\left(\frac{R}{z}\right)\right)$$

The numerical aperture (NA) of the MDL design varies with z from the diffraction limit, and R is the radius of the MDL. Here, $\lambda / \max(NA)^2$ is the diffraction-limited depth of focus (DOF) for the MDL design [56]. More details on design methodology were reported previously in [53]. Eventually, the focusing efficiency of MDL is maximized by selecting the distribution of the heights of its individual rings [51, 56]. The present MDL design consists of concentric rings with a width of 7 μm and a height that varies from 0 to 1.7 μm [54]. The present MDL was designed for a wavelength of 1064 nm and possesses a working distance of ~20 cm (corresponding to its focal length). Additionally, the focus length is extended by 5 cm in space. The multi-level diffractive lens (MDL) pattern depicted in Fig. 1(a) was etched onto a polymer photoresist film (S1813, Microchem) mounted on a glass wafer employing the greyscale lithography technique (DWL 66+, Heidelberg Instruments Mikrotechnik GmbH) with an aperture diameter of 3 mm. The corresponding phase distribution is illustrated in Fig. 1(b). The reason the polymer photoresist was chosen is that this material exhibits high transmission across a broad range of wavelengths, and the overall thickness of the MDL is primarily determined by the substrate thickness considered in the fabrication process [51]. However, the complete details of the fabrication with a similar MDL are reported in [53]. Additionally, the applied fabrication technique is relatively simple and inexpensive compared to cutting-edge e-beam lithography with the subsequent lift-off technique, collectively making our experimental approach cost-effective. Previously, the MDL device with similar functionalities has been developed and shown to operate over a broad wavelength range [50, 57]. Furthermore, the MDL was utilised to generate optical vortex needle beams, focusing an input optical vortex at a desired working distance while simultaneously increasing the depth of focus (DOF) to several times [54]. Recently, the MDL was employed to create longer plasma channels for strongly focused femtosecond laser pulses that are unattainable with a traditional focusing lens [55]. These studies demonstrate that MDL is highly advantageous in applications and highly effective for focusing the laser tightly, while also extending the focus over a longer range.

To verify the generation of Gaussian beams with an extended depth of focus, we illuminate the MDL with a Gaussian beam of 3 mm spot size, an input beam of 3 mm in size (measured using the 1/e² method), whose 2D and 1D intensity distributions in the beam cross-section are shown in the respective Figs. 1(c) and 1(d). The propagation of the focused Gaussian beam is shown in Fig. 1(e). As evident, we have obtained a high-quality, focused Gaussian beam that remains focused and whose shape and size remain unchanged over a finite distance (extended depth of focus), indicating its non-diffracting nature in space. The focusing of a Gaussian beam through MDL is compared to that of a conventional lens ($f$ = 20 cm). As is evident, we have obtained a focused Gaussian beam whose focus is extended to a very short distance; subsequently, the beam diffracts or diverges, as illustrated in Fig. 1(f). These results reveal that the flat diffractive focusing lens serves as a brilliant alternative to conventional focusing lenses employed in optical experiments, enabling the generation of high-quality focused light beams with an extended depth of focus —a highly desirable property in several applications. In this perspective, to explore the exciting performance and impact of flat optics that impart a multilevel diffractive phase on propagating light beams, we investigated the propagation of the zero-order Bessel laser beam through a multi-level diffractive lens (MDL), as reported in Sections III and IV, including experimental findings.

## III. METHODS AND EXPERIMENTAL ARRANGEMENT

### A. Operating principle

Upon illumination with a Gaussian beam, the binary axicon produces Bessel beams characterised by varying sets of wavevectors that are superimposed and interfere during propagation, culminating in the formation of an optical bottle exhibiting partial characteristics of dark and bright cores that are scarcely distinguishable (Fig. 2(a)). Incorporating the MDL after the binary axicon results in a high-quality, micron-scale optical bottle beam exhibiting sharp intensity features (Fig. 2(b)). Nevertheless, the far-field intensity profile of the Bessel beam generated through a conventional lens differs from that produced with the MDL, as visualised in the schematics.

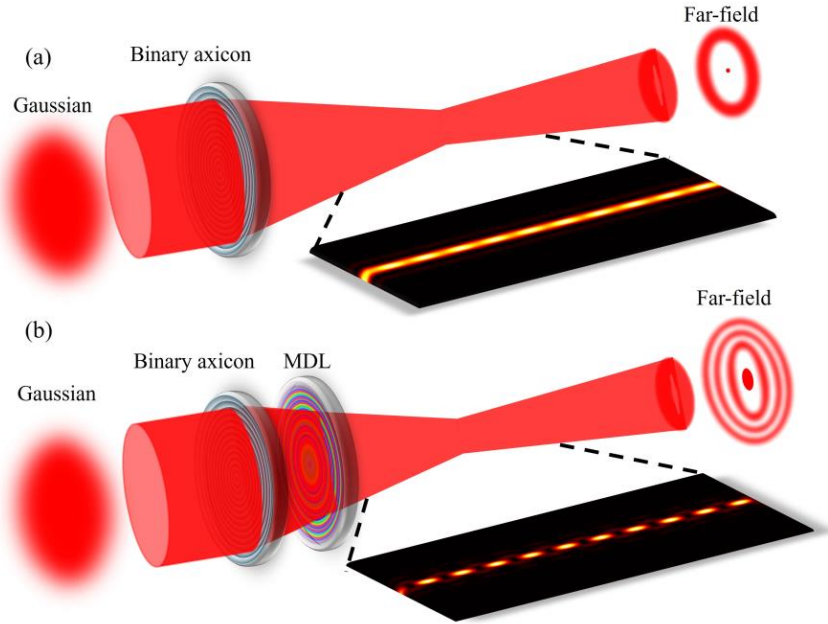

**Fig. 2.** Schematics of methods generating a Bessel optical bottle beam (a) with a binary axicon (b) with a binary axicon and a multi-level diffractive lens (MDL).

### B. Experimental setup

A linearly polarised Gaussian laser beam that passes through a beam-expanding setup built with $L_1$ and $L_2$ lenses successfully produces a Gaussian beam with a 10X magnification with collimation, which directs the expanded beam to illuminate a binary axicon having an aperture size of ~30 mm. It consists of Bessel zones with 30 $\mu$m periodicity. As a result, the axicon shapes the input Gaussian beam into an output Bessel beam. Furthermore, the generated Bessel beam directs towards the MDL and tightly focuses the laser beam in space. Subsequently, the focus (marked as the $z$ = 0 plane) extends over a longer distance in free space, as shown in Fig. 3(a). The CCD camera displaces in the $z$-direction to collect the resultant intensity distributions at different distances along the beam propagation. The experimental photo shown in Fig. 3(b) displays the original experimental setup, including optical elements and the camera, all of which are aligned to generate a high-quality optical bottle beam for characterizing intensity distributions corresponding to low- and high-intensity regions of the non-diffracting bottle beam. For precise alignment, we mount the diffractive optical elements on 3D translation stages with a minimum step size of 10 microns.

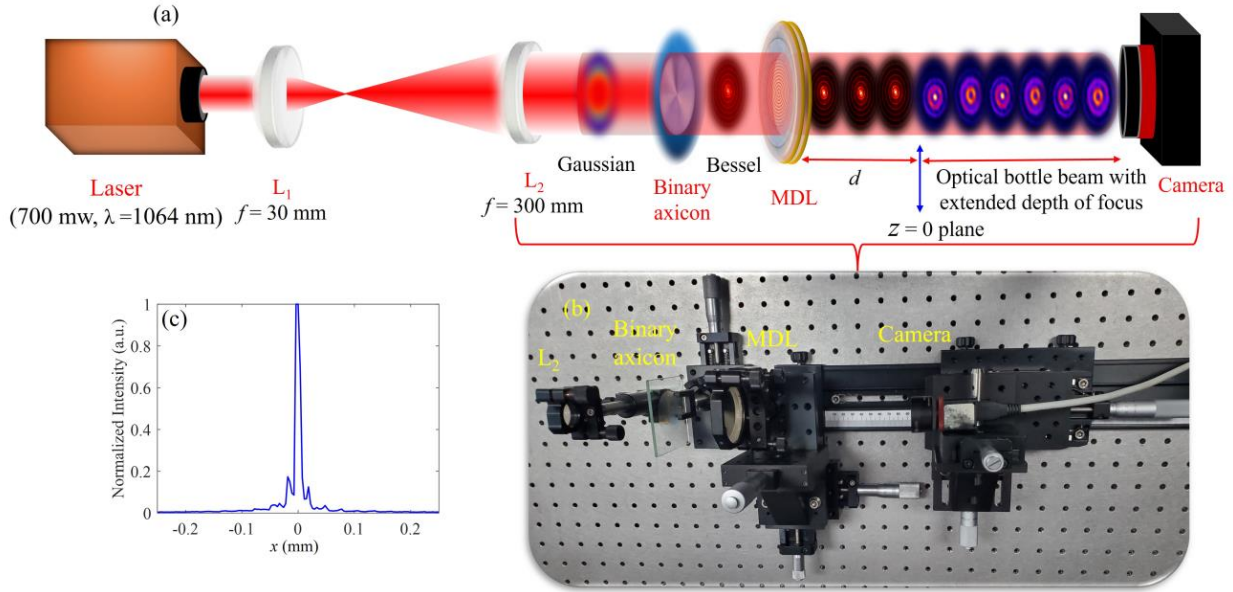

**Fig. 3.** (a) The experimental schematic showing the formation of the optical Bessel bottle beam, where the focus or working distance of the MDL is, and the $z=0$ plane represents the point located in free space at a distance from the MDL where the dynamic beam shaping of the input beam has taken place with an extended depth of focus property. In this context, the MDL is illuminated by the input beam, which is a Bessel beam generated from the binary axicon mounted before the MDL at a distance of ~55 mm. In this configuration, a binary axicon generates a non-diffracting Bessel beam when a linearly polarised laser beam with a Gaussian distribution incident at a wavelength of 1064 nm is expanded with a pair of plano-convex lenses $L_1$ and $L_2$ whose focal lengths are $f = 30$ mm and $f = 300$ mm, respectively. (b) A photo of an experimental setup, in which an axicon, MDL, and a CCD camera are mounted on XYZ translation stages for precise alignment to the beam direction that leads to detecting intensity distributions at various distances in free space before and after the $z = 0$ plane marked in the schematic shown in Fig. 3(a). (c) Intensity profile of the Bessel beam with flattened sidelobes generated from the binary axicon fed as input to the MDL.

The diffractive phase profile of the binary axicon imprinted into the UVFS substrate using lithography showcases leading characteristics, including a high damage threshold, a diffraction efficiency of up to 95%, and a light transmission of 90% at $\lambda = 1064$ nm, making it a suitable product for generating a Bessel beam with a high-intensity central lobe and flattened sidelobes realizing beneficial results in laser material processing (Fig. 3(c)). Specifically, this approach simplifies the experimental scheme into a compact form by utilizing a binary axicon and MDL presented diffractive optical elements, making it economical and commercially useful to generate high-quality, high-energy-density Bessel beam bottles.

## C. Propagation of a Bessel beam with intensity voids

To verify the propagation dynamics of the optical Bessel beam originating from the binary axicon without a focusing lens (MDL), we measured the transverse intensity distribution of the beam along the propagation direction, as shown in Fig. 4. The representative intensity distributions of the Bessel beams obtained at various distances after Gaussian beam illuminated on binary axicon are illustrated in Fig. 4(a). The corresponding transverse cross-sectional intensity profiles of the Bessel beam are also included in Fig. 4(a). Note that the output quality of the propagating Bessel beam decreases with increased distance in free space, clearly depicted in Figure 4(a).

Furthermore, we present the longitudinal intensity distribution of the Bessel beam generated by a binary axicon in Fig. 4(b). Although we scanned the intensity distribution over a propagation distance ($z$) of 100 mm, the intensity profile of the beam along 35 mm from the plane, where the optical bottle beam signature appears, is highlighted in Fig. 4(b).

As is clear, the generated beam shows regions of low and high intensity that emerge during propagation. The low-intensity regions along the propagation axis have an intensity similar to that of the high-intensity regions since this optical bottle structure in the Bessel beam is produced from the interference of beams with two different sets of wave vectors, $k_1$ and $k_2$. As the beam propagates in free space, the interference among these wavevectors varies, resulting in an optical bottle beam exhibiting partial characteristics of dark and bright cores that are hardly distinguishable due to oscillating intensities over regions along the beam propagation (Fig. 4(b)).

After the binary axicon, the Bessel beam established at $z = 50$ and $55$ mm exhibits superior quality in comparison to those obtained at subsequent distances in free space. Therefore, a high-quality Bessel beam was selected as the input to illuminate the MDL, thereby demonstrating its optimal performance during experimental measurements. Accordingly, the distance between the axicon and the MDL is adjusted to 55 mm, ensuring that the Bessel beam of appropriate dimensions serves as an effective input to fully exploit the potential of the MDL phase. This multi-level diffractive phase dynamically influences the intensity distribution of Bessel beam propagation from the $z = 0$ plane (focus), and this effect persists over a longer range.

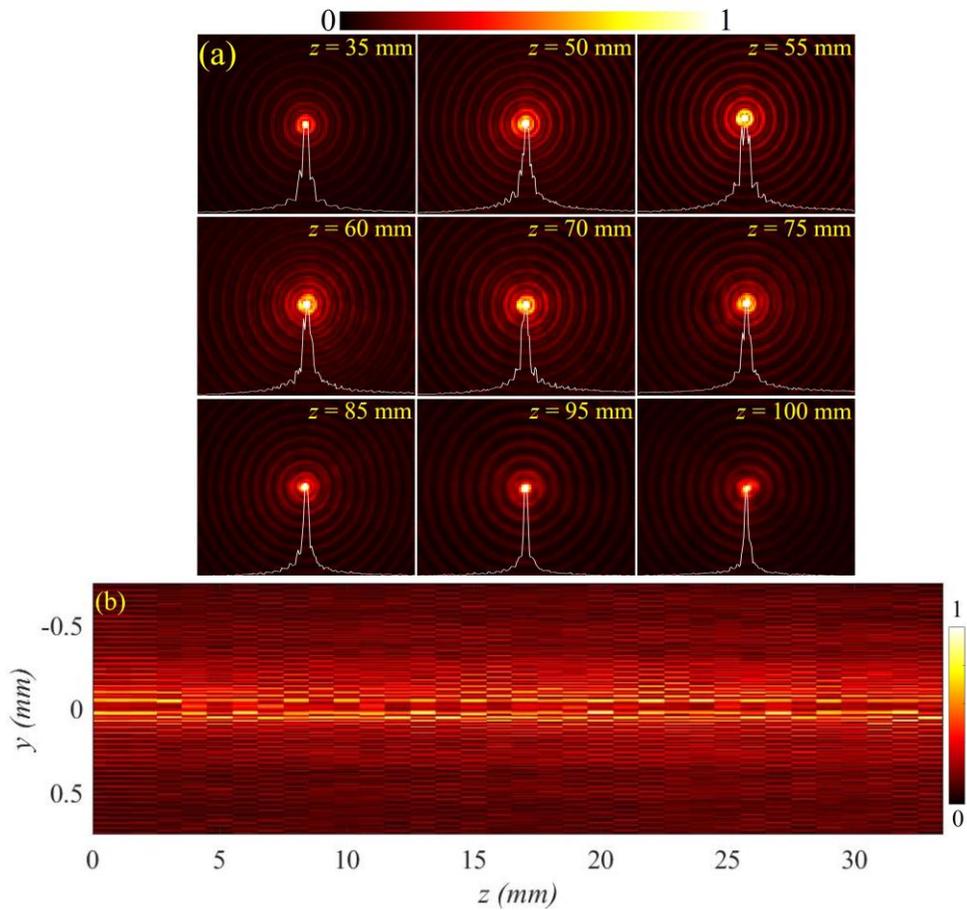

**Fig. 4.** Experimental intensity distributions of the Bessel beam originated from the binary axicon obtained at various distances during free space propagation, and corresponding cross-sectional intensity profiles have also been included. (a) Transverse intensity distribution of a Bessel beam (*xy*-plane) at different propagation distances. (b) Longitudinal intensity distribution of a Bessel beam (*yz*-plane), where low- and high-intensity regions are seen along the beam propagation function of distance (*z*).

## IV. GENERATING HIGH-QUALITY BESSEL BOTTLE BEAMS WITH EXTENDED DEPTH OF FOCUS

To verify the generation of high-quality Bessel bottle beams, the MDL was placed after the binary axicon at a distance of 55 mm, yielding the transverse intensity profile of the beam along its propagation direction. The results are shown in Fig. 5. As is evident in Figs. 5 (a) & (b), the longitudinal intensity distributions for the generated optical Bessel bottle beam are shown in the $xz$ and $yz$ planes. The corresponding cross-sectional profiles are shown as insets in Figs. 5(a) & (b). These representative intensity distributions effectively demonstrate the dynamic effect of the MDL diffractive phase on the Bessel beam as it passes through it. The measured focal distance of the MDL for the illuminated Bessel beam is set as a $z = 0$ plane (marked in Fig. 3(a)), and the MDL working distance for the input Bessel beam is detected as $d \approx 19.5$ cm. The MDL phase facilitates strong interference of beams with different sets of wavevectors while fine focusing, and this interference effect remains stronger and consistent over an extended depth of focus. As a result, a series of zero-intensity and bright-intensity regions emerge alternately, forming a micron-sized high-quality optical Bessel beam that remains shape-invariant and non-diffracting over a long distance of ~75 mm, as illustrated in Figure 5. Zero-intensity (dark) and high-intensity (neck) regions appear alternately with sharp features, spanning a non-diffracting length of the optical bottle beam, and are separated by a variable period ranging from 0.2 to 1.36 mm. Interestingly, no divergence was observed for the generated Bessel bottle beam within a resolution of 3.69 μm. This exceptional non-diffracting property over a long range, as demonstrated by the optical bottle beam, offers potential benefits for optical focusing systems employed in micromachining and microfabrication.

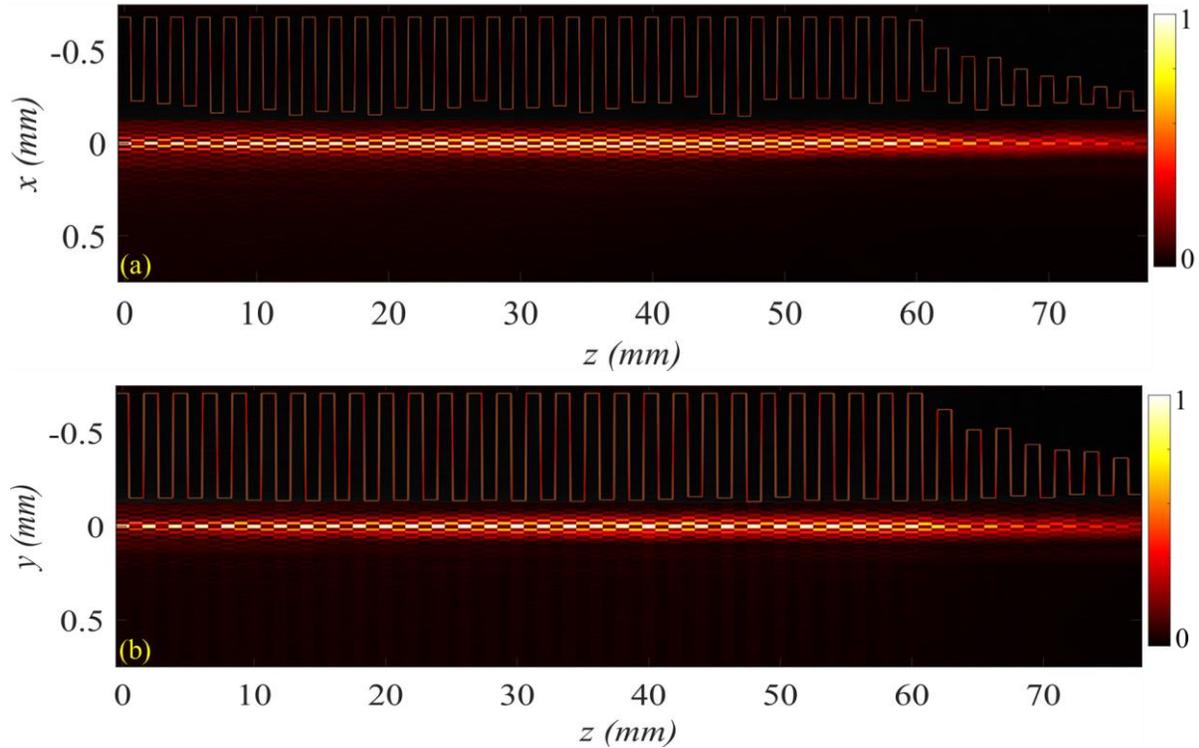

**Fig. 5.** Evolution of optical bottles in the experimentally generated Bessel profile by the combination of a binary axicon and multi-level diffractive lens (MDL). Longitudinal intensity distribution of Bessel bottle beams in the (a) $xz$ - plane and (b) $yz$ - plane.

Additionally, the individual intensity distributions at different planes, along with their corresponding cross-sectional intensity profiles, are shown in Fig. 6. These results were recorded using a CCD camera positioned at various distances in the $z$-direction as the Bessel optical beam propagated in free space. During the observation of low-intensity (dark) regions, the variation in period remains minimal for initial propagation distances in comparison to high-intensity (neck) regions, and it increases as the propagation distance increases. The MDL is highly effective in forming an optical bottle beam with sharply defined dark and neck regions. This strong effect extends over a distance of ~ 60 mm in free space (from $z \sim 0$ to $z \sim 60$ mm), where the output quality and intensity of the beam profiles are relatively the same, as shown in Figure 6. Upon further propagation of the beam, the intensity of the beam profiles decreases, and note that the spatial distribution of dark and neck regions in the optical bottle beam remains invariant to propagation, regardless of the decrease in the intensity. The ring structure observed in beam profiles (Fig. 6) around the centre relies on the input Bessel beam alignment to the MDL centre, but not the illuminating area of the MDL surface. As the MDL is a new class of diffractive lens with unique focusing features, most of the focused light energy is tightly concentrated in the central part of the intensity profile. The remaining small portion of light energy is distributed over the ring structure or the sidelobes region, which proves the Bessel bottle beam is non-diffracting and non-divergent over a long propagation distance.

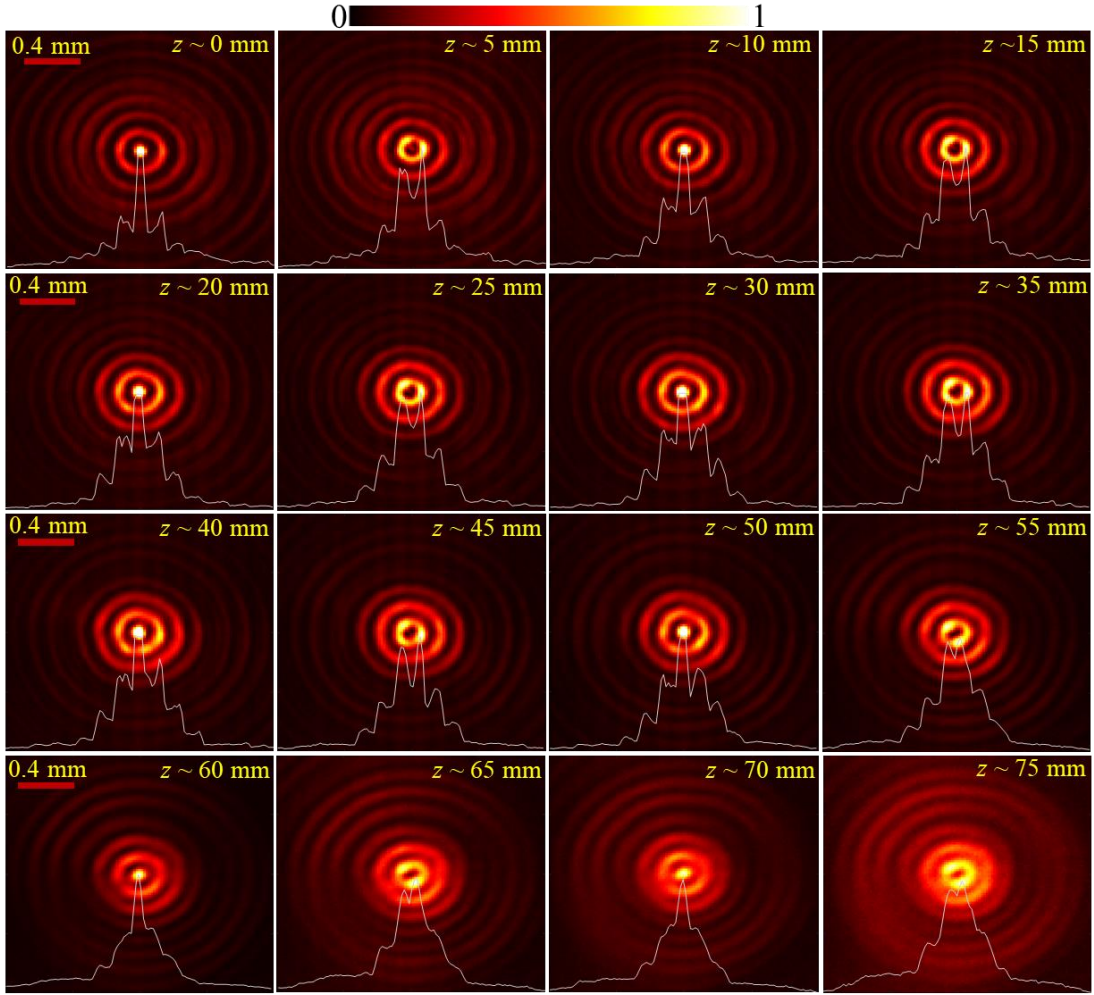

**Fig. 6.** Experimental intensity distributions of Bessel bottle beams generated by a multi-level diffractive lens and corresponding to dark (zero-intensity centre) and neck (high-intensity centre) regions alternatively appear as a function of propagation distance ($z$) in mm. The corresponding transverse cross-sectional intensity profiles at different propagation distances are included here.

Furthermore, we have evaluated the normalized peak intensity for transverse intensity profiles corresponding to neck and dark regions detected at various distances $z$ (mm) along the spatial intensity distribution of the Bessel bottle beam, and also evaluated the focusing efficiency (η) as shown in Figs. 7(a), 7(b), and 7(c), respectively. The ratio of light power encircled inside the area of spot diameter is equivalent to three times FWHM (full width at half maximum) to the incident light power, which is determined as focusing efficiency (η) [54, 56]. As is evident, the normalised peak intensity remains maximum for the optical bottle beam up to ~60 mm distance, and thereafter, it monotonically decreases (Fig. 7(a)) for axial profiles corresponding to the neck regions, and also decreases for the dark regions (Fig. 7(b)). Further, the optical bottle beam shows good focusing efficiency (η) over long propagation distances. It remains almost constant for distances up to ~35 mm, i.e., it varies between 79% to 78%, and it decays slowly approaching 74.5% at ~60 mm, as shown in Fig. 6(c). After that, the focusing efficiency declines further, reaching 31.5% for longer propagation distances.

In the current investigations, we utilised an MDL with a clear aperture of 3 mm. It is important to note that when the MDL is illuminated by an input light beam that is either smaller or larger than 3 mm, the extended depth of focus of the output light beam and the corresponding focusing efficiency (η) are affected by significant factors. A comprehensive investigation into such scenarios has already been documented in reference [54]. Consequently, an input Bessel beam approximately 3 mm in size, generated by the binary axicon, was employed to illuminate the MDL. This input Bessel beam features a complex ring structure, with the intensity residing in lower-order sidelobes being relatively weak or small, while the higher-order sidelobes exhibit zero intensity. Furthermore, the extended depth of focus and the working distance ($z = 0$ plane) can be adjusted by modifying the parameters for which the MDL was designed, thereby enabling it to operate over a broad spectral range [50, 57].

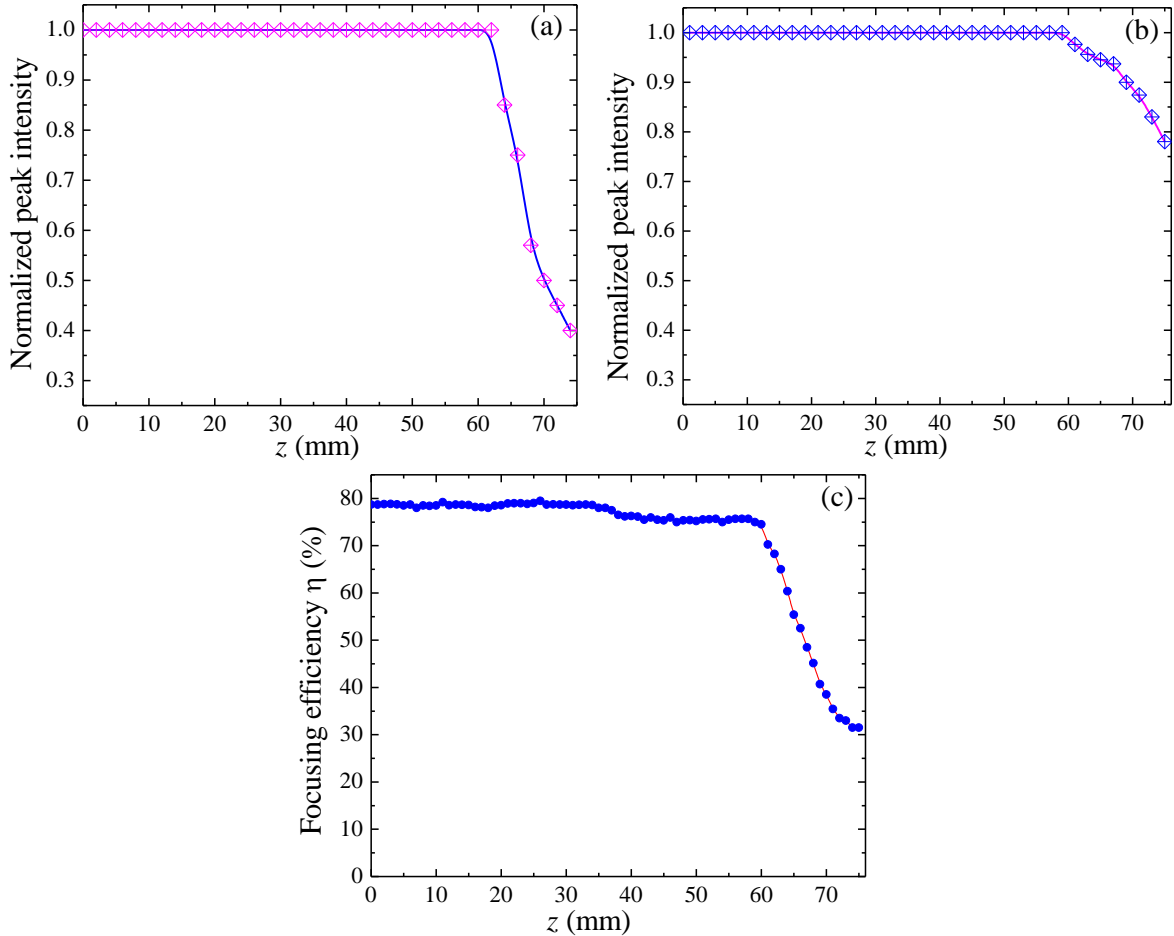

**Fig. 7.** Normalized peak intensity of Bessel bottle beams generated by a multi-level diffractive lens, as a function of $z$ (a) for high-intensity regions, and (b) for low-intensity regions. (c) focusing efficiency (η) of the propagating Bessel bottle beam as a function of $z$. The input size of the Bessel beam is considered as ~3 mm.

Interestingly, the MDL has shown a promising influence on the incident Bessel beam before approaching its focus position. In this context, we have recorded experimental intensity distributions at various distances before approaching the focus along the beam propagation from $z$~15 cm to $z$~19 cm in free space, as shown in Fig. 8.

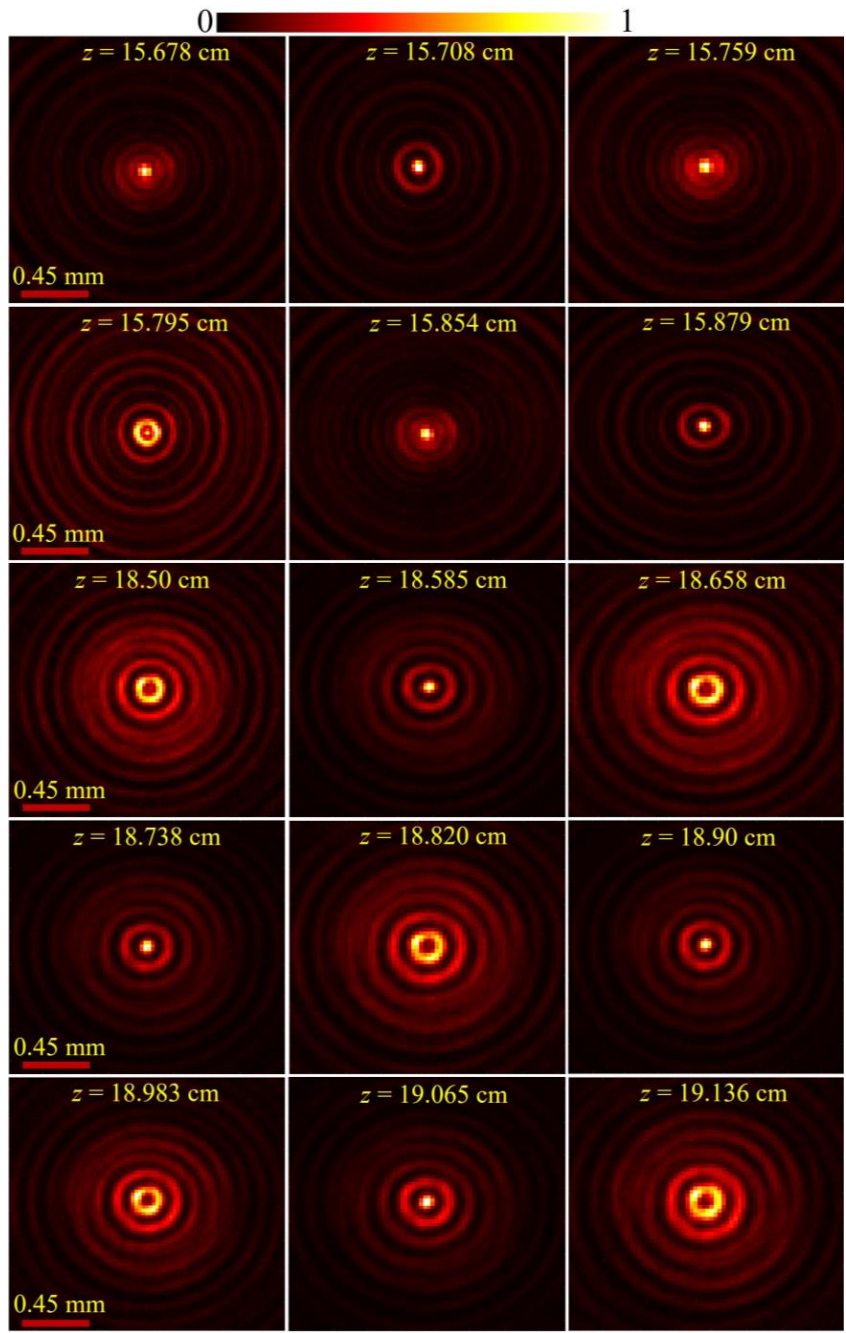

**Fig. 8.** Evolution of experimental intensity distributions is employed to observe the transformation of a propagating Bessel beam via a multi-level diffractive lens before reaching its focus ($z = 0$, focal plane) along the working distance (*d*).

Initially, we observed no interference of beams of wavevectors characterised by the absence of an optical bottle structure in the beam propagation dynamics. As is evident, experimental intensity distributions are shown at different distances $z$ = 15.678 cm, 15.708 cm, 15. 759 cm, 15.795 cm, 15.854 cm, and 15.879 cm. The dynamic effect of the MDL diffractive phase on the Bessel beam becomes apparent at $z$ = 18.5 cm, where the interference of the beams occurs. Subsequently, low-intensity and high-intensity regions parade alternatively, leading to the observation of a Bessel bottle beam structure. This promising effect is illustrated in the results shown in Fig. 8, which span a propagation distance range of $z$ = 18.50 cm to $z$ = 19.136 cm. These successive intensity profiles were observed alternately, separated by a variable period ranging from 0.6 mm to 0.83 mm, until they approached the $z$ = 0 plane, which appeared over a working distance of $d \approx$ 19.5 cm from the MDL position marked in the experimental schematic (Fig. 3(a)).

The MDL is dedicated to extending the depth of focus of incident light beams that are not attainable with a traditional focus lens of the same focal length ($f \sim$ 20 cm). A traditional lens typically performs a Fourier transformation when focusing the Bessel beam at the focal plane. In contrast, the MDL offers a unique ability by focusing structured light beams, highlighting a novel approach to focusing Bessel beams through its dynamic phase, which generates a distinct beam profile in the far-field plane. With MDL, the far-field intensity distribution of the Bessel beam differs from that obtained with a traditional focusing lens, offering an interesting perspective on beam focusing, as depicted in Fig. 9, along with corresponding cross-sectional intensity profiles. The transverse intensity distribution formed by MDL is shown in Fig. 9(a) and was detected at a propagation distance of ~35 cm (far-field plane) from the MDL position, as indicated in the experimental schematic (Fig. 3(a)). The intensity distribution shown in Fig. 9(a) consists of a uniform ring structure around the central bright spot. The intensity distributed over the ring region is relatively low, i.e., side lobes are found with low intensities. The resultant beam profile at the far-field plane retains non-diffracting features, offering potential benefits in light-matter interaction experiments and applications. By contrast, the far-field intensity profile in Fig. 9(b) is generated by a traditional lens at the focal length ($f$ = 20 cm), representing the Fourier transformation of the input Bessel beam, which is an annular beam with a central spot and annular rings of different intensities. Although the far-field of a traditional lens offers advantages, the far-field formed by the MDL offers more potential advantages over a traditional focusing lens.

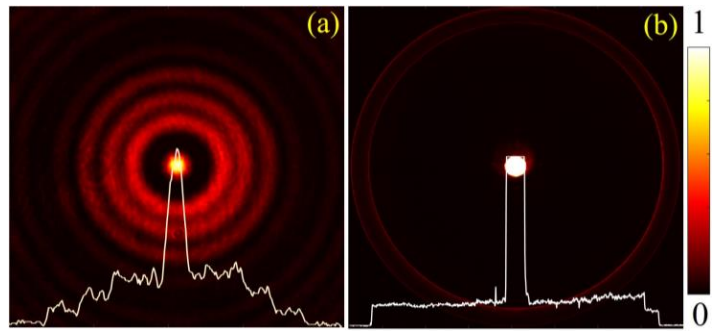

**Fig. 9.** The far-field intensity distributions obtained by different focusing lenses are as follows: (a) the far-field of a Bessel beam generated with the multi-level diffractive lens over a propagation distance of $z$~35 cm, (b) the far-field generated with the traditional focusing lens over a focal distance of $f$ = 20 cm.

## V. DISCUSSION

The diffractive optical elements, including a binary axicon and a multi-level diffractive lens, possess a notably high damage threshold, making them suitable for generating ultra-short Bessel laser pulses with pulse energies on the order of millijoules. The wavelength of these high-energy laser pulses can be tuned to the ultraviolet spectrum via nonlinear wave mixing in solids and gases, thereby extending the applications of optical bottle beams to non-lasing wavelengths [8, 12, 35]. The inherent self-imaging, non-diffracting, and self-reconstructing properties of Bessel bottle beams can be exploited for concurrent dark and bright trapping in parallel transverse planes [58]. The reduced sidelobes of Bessel beams produced by binary axicons facilitate high-quality volumetric imaging [59] and light-sheet imaging [60], even under conditions of very low signal-to-noise ratio, without necessitating nonlinear interactions. The needle-like beam intensity distribution with minimised sidelobes derived from binary axicons presents an excellent option for material processing, surpassing conventional Bessel beams [61-65]. Furthermore, three-dimensional non-diffracting optical potentials generated in Bessel beams can be further modulated using a multi-level diffractive lens (MDL) [66]. The Gaussian beam, when pumped through an MDL, offers an alternative method with an added advantage over hollow Gaussian and annular beams, providing an extended depth of focus and thus enhancing high-resolution bio-imaging applications [67-68]. Additionally, structured modes produced by a versatile structured laser source can be converted into unique 3D structures using the MDL presented in this work.

## VI. CONCLUSIONS

We have developed and demonstrated a novel method for generating a non-diffracting, micron-sized optical bottle beam, providing a simple, compact, and cost-effective experimental approach. We demonstrated that a Bessel beam propagates through MDL, producing a tightly focused output beam that consists of a series of low-intensity regions surrounded by high-intensity regions, which alternately appear in free space and eventually fuse to form a Bessel optical bottle beam. Notably, the dynamic focusing of the optical bottle beam effectively starts at the MDL working distance of $d \approx 19.5$ cm and remains propagation-invariant over its extended depth of focus. The resultant bottle beam retains its shape and size over a long distance (>5 cm), demonstrating its nondiffracting property. Furthermore, we have shown the influence of the MDL on the Bessel beam at various distances, both before and after the $z = 0$ plane, along the propagation direction. Thus, MDL has proven to be a highly promising candidate for generating high-quality, nondiffracting Bessel beam bottles. Additionally, the far-field intensity distribution produced by the MDL with the input Bessel beam offers more valuable insights for light-matter interactions compared to the far-field intensity distribution generated by a traditional focusing lens of the same focal length ($f = 20$ cm). All our experimental results hold significant potential for applications such as high-contrast biological imaging, micromachining, microfabrication, optical tweezers, particle manipulation, and structured high-harmonic generation. Moreover, we plan to advance our research by creating high-power, high-quality 3D Bessel optical bottle beams across a broad spectral range, including the UV and XUV regimes.

**FUNDING**

The authors acknowledge support in the form of KAKENHI Grants-in-Aid (Nos. JP22H05131, JP22H05138, JP23H00270) from the Japan Society for the Promotion of Science (JSPS) and from the Core Research for Evolutional Science and Technology program (No. JPMJCR1903) of the Japan Science and Technology Agency (JST). V. P. gratefully acknowledges funding support from Science and Engineering Research Board (CRG/2021/248003060). T. M. and R. M. gratefully acknowledge funding from United States Office of Naval Research (N00014-22-1-2014).

**ACKNOWLEDGEMENT**

The corresponding author expresses gratitude to Hasan Yılmaz (Institute of Materials Science and Nanotechnology, National Nanotechnology Research Center (UNAM), Bilkent University, Turkey) for fruitful discussions, and to Vasu Dev (NUS, Singapore) for helpful remarks during experiments.

**AUTHOR DECLARATIONS**

**Conflict of Interest**

The authors declare no conflicts of disclosure.

**DATA AVAILABILITY STATEMENT**

The data and datasets underlying the results presented in this paper are available from the corresponding author upon reasonable request.